# Narrowly distributed crystal orientation in biomineral vaterite

Boaz Pokroy[1], Lee Kabalah-Amitai[1], Iryna Polishchuk[1], Ross T. DeVol[2], Adam Z. Blonsky[2], Chang-Yu Sun[2], Matthew A. Marcus[3], Andreas Scholl[3], Pupa U.P.A. Gilbert[2,4,5,*,#]

[1] Department of Materials Science & Engineering, and the Russell Berrie Nanotechnology Institute, Technion - Israel Institute of Technology, Haifa 32000, Israel.
[2] Department of Physics, University of Wisconsin–Madison, 1150 University Avenue, Madison, WI 53706, USA.
[3] Advanced Light Source, Lawrence Berkeley National Laboratory, 1 Cyclotron Road, Berkeley, CA 94720, USA.
[4] Department of Chemistry, University of Wisconsin–Madison, 1101 University Avenue, Madison, WI 53706, USA.
[5] Radcliffe Institute for Advanced Study, Harvard University, 8 Garden Street, Cambridge, MA 02138.

*Previously publishing as Gelsomina De Stasio.
#Corresponding author: pupa@physics.wisc.edu

## Abstract

Biominerals formed by animals provide skeletal support, and many other functions. They were previously shown to grow by aggregation of amorphous nanoparticles, but never to grow ion-by-ion from solution, which is a common growth mechanism for abiotic crystals. We analyze vaterite ($CaCO_3$) multi-crystalline spicules from the solitary tunicate *Herdmania momus*, with Polarization-dependent Imaging Contrast (PIC)-mapping, scanning and aberration-corrected transmission electron microscopies. The first fully-quantitative PIC-mapping data, presented here, measured 0°-30° angle spreads between immediately adjacent crystals. Such narrowly distributed crystal orientations demonstrate that crystallinity does not propagate from one crystal to another (0° angle spreads), nor that new crystals with random orientation (90°) nucleate. There are no organic layers at the interface between crystals, hence a new, unknown growth mechanism must be invoked, with crystal nucleation constrained within 30°. Two observations are consistent with crystal growth from solution: vaterite microcrystals express crystal faces, and are smooth at the nanoscale after cryo-fracture.

The observation of 30° angle spreads, lack of interfacial organic layers, and smooth fracture figures broadens the range of known biomineralization mechanisms and may inspire novel synthetic crystal growth strategies. Spherulitic growth from solution is one possible mechanism consistent with all these observations.

# Introduction

Biominerals are polycrystalline minerals formed by living organisms, with a multitude of functions, including skeletal support [1], locomotion, biting [2], mastication [3], attack and defense tools [4], gravity and magnetic field sensing [5,6], and many others [7]. Biominerals nearly always include intra- and inter-crystalline organic molecules [8], even when pathological mineralization occurs [9,10], and at the end of their diverse formation mechanisms they result in hard and tough tissues with varying degrees of crystal co-orientation: from the single-crystalline sea urchin spicules and spines to randomly oriented poly-crystalline aragonite in the outer part of *Nautilus* shells. Bone, teeth, various mollusc and brachiopod shell structures, all have intermediate crystal orientation angle spreads [11-21]. Rarely do organisms utilize vaterite ($CaCO_3$) as their mineral components [7,22], possibly because vaterite is more soluble and less stable than calcite and aragonite (both also $CaCO_3$). One such organism is the sea squirt *Herdmania momus* [23], a tunicate that forms vaterite spicules in its tunic and body presumably for stiffening these tissues, while maintaining a flexible structure. Its spicules exhibit a unique morphology: a series of pointy crystals arranged in a "crown of thorns" motif, which helically surround elongated core fibers. Each thorn is a larger, higher-quality single crystal than any geologic or synthetic vaterite ever observed. These thorns were therefore used recently to reveal the double-structure of vaterite: electron phase contrast imaging along the *c*-axis of a single thorn revealed a major hexagonal structure identical to the one described by Kamhi [24] (space group of P63/mmc with a= 4.13A and c=8.49A), whereas the other minor structure has symmetry still unknown and larger crystal lattice spacing [25,26].

In this work we investigate entire vaterite spicules from the same animal, their crystal orientations, and their formation mechanism, by using Polarization-dependent Imaging Contrast (PIC)-mapping [27-34], a mode of PhotoEmission Electron spectroMicroscopy (PEEM) [35], Scanning and aberration corrected Transmission Electron Microscopies (SEM and TEM). The TEM data also include nanobeam linescan electron diffraction.

# Results

### Crystal orientation measurements

In Figure 1 we present a montage of all the vaterite spicules analyzed in this work, along with synthetic vaterite. In synthetic vaterite we observe micron-size domains, each of which includes many co-oriented nanoparticles, whereas domains are randomly oriented with respect to one another. Each biogenic vaterite spicule, instead, shows a small angle spread of crystal orientations, evidenced by similar colors.

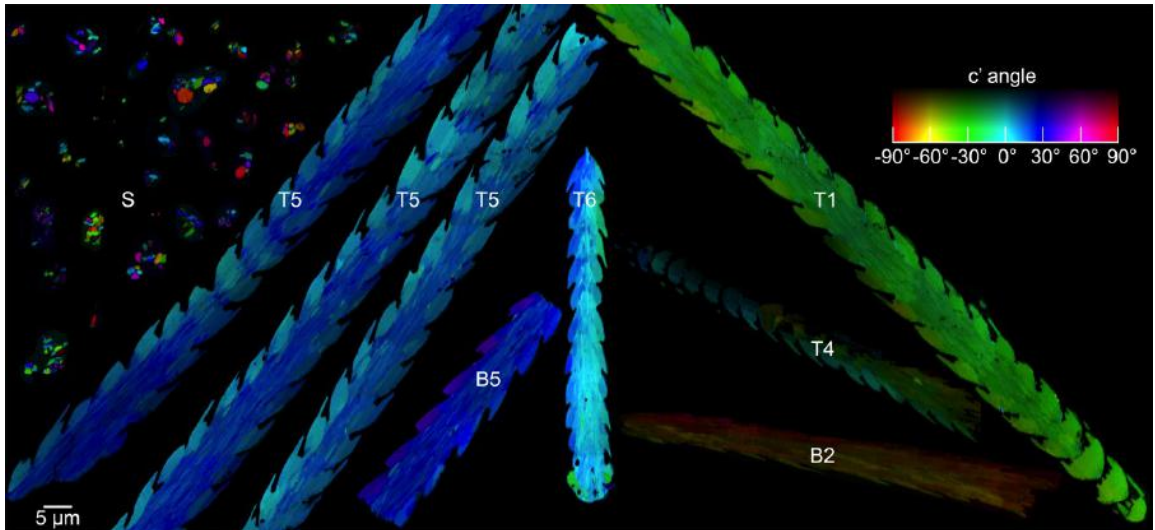

*Figure 1: Composition of PIC-maps of vaterite tunic (T) and body (B) spicules from H. momus in polished cross-section. Color and brightness quantitatively represent the orientation of vaterite crystals, and demonstrate that each spicule is multi-crystalline, with small-angle branching of crystallites as shown by different but nearby colors in each spicule. Thorns, protruding from the core of each spicule, are the largest singly oriented crystals. Synthetic vaterite crystalline domains (S) are randomly oriented. Spicule T5 is segmented for efficient space use. The color bar displays in different hues different angles between the vertical and the c'-axis (projection of the vaterite c-axis onto the polarization plane, which is in turn perpendicular to the x-ray beam, and is tilted by 60° around the vertical with respect to the image plane shown here). A crystal with vertical c- or c'-axis is cyan, a horizontal one is red. Brightness displays how far off-plane the c-axis is oriented. Dark crystals, e.g. in spicules T4 and B2, have their c-axes nearly normal to the polarization plane, bright crystals have their c-axes in the polarization plane, as in spicule T6.*

In a PIC-map different colors correspond to different crystal orientations, measured based on linear dichroism [36], an x-ray effect that makes π* peaks in carbon and oxygen absorption spectra vary in intensity depending on the orientation of π–bonded carbonate groups. When π orbitals are parallel/perpendicular to the linear polarization of the illuminating x-ray beam, the peak has maximum/minimum intensity, respectively, because the dipole interaction is maximum/minimum. In a PIC-map this orientation sensitivity is exploited to visually display crystal orientations.

In Figure 2 we show the tip of spicule T1 physically rotated to be imaged in three different positions. They therefore provide a complete 3D quantitative description of the vaterite c-axis orientation, rather than its projection onto a 2D plane. This is the first fully quantitative measurement of the arrangement of single crystals in their pristine crystal orientation pattern, obtained with 20 nm resolution. Figure S1 shows a level-enhanced, non-quantitative version of Figure 2, to display vaterite crystals, even in the positions in which one can hardly see them in Figure 2.

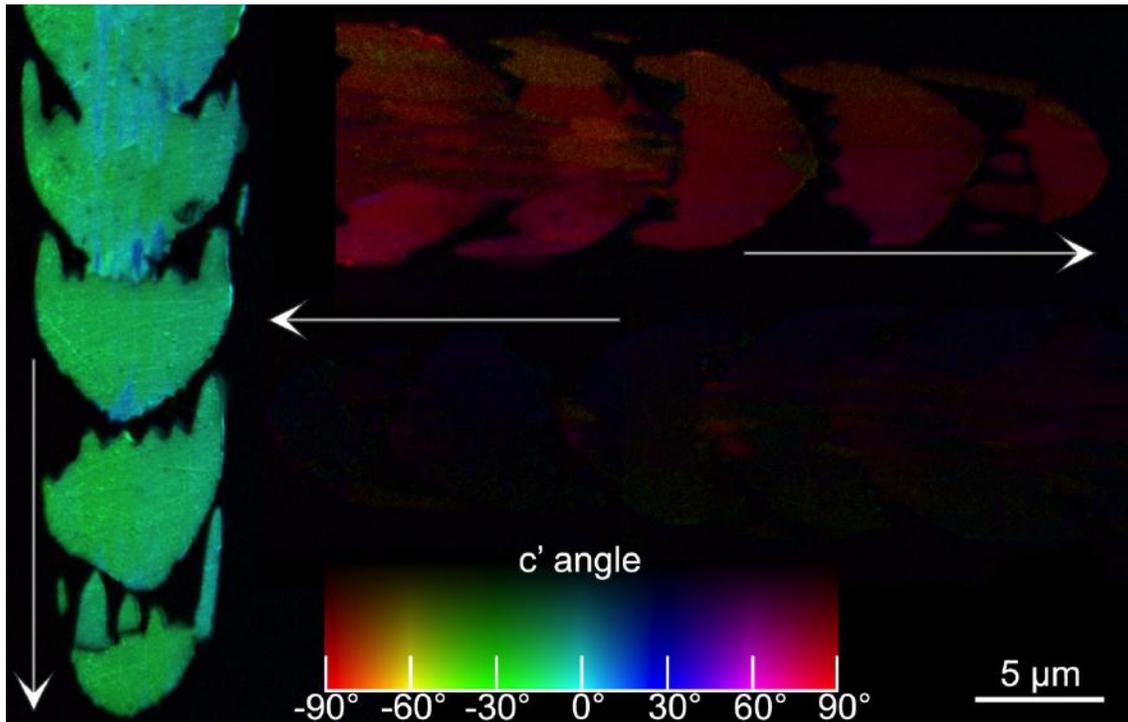

*Figure 2: PIC-maps of the same spicule (T1 in Figure 1) imaged after physically rotating the sample. In all 3 PIC-maps the arrows point towards the tip of the spicule. In the vertical position the crystals are blue-cyan-green, indicating vertical or nearly vertical c'- and c-axes. When the spicule is rotated horizontally, therefore, the colors become orange-red-magenta to indicate a general horizontal direction of c'-axes. In the horizontal positions the crystals are darker as expected, because the image plane is 60° from the polarization plane. In addition, the two horizontal positions are different from one another: the left-pointing spicule is almost completely black, indicating that the c-axes of its crystals are not in the image plane but 30° off-plane, perpendicular to the polarization plane, that is, they point straight into the incident x-ray beam. See Figure S1 for a brighter version of this figure.*

In Table 1 we present the measured angular distances between the *c'*-axes and *c*-axes in pairs of adjacent crystals. Crystalline *c*-axes may be oriented anywhere in 3D, and their 3D orientation is measured, whereas *c'*-axes are projected onto the 2D polarization plane, which is 60° from the image plane, rotated around the vertical in the image, which is also the laboratory's vertical. The measured numbers are in excellent agreement with one another, because for this spicule all crystalline *c*-axes are nearly in the polarization plane, which is the most favorable case. In the general case, if *c*-axes are farther off-plane, the 3D angular distance Δ*c* is perfectly quantitative, but the 2D Δ*c'* could be either larger or smaller depending on the projection angle onto the polarization plane. Notice that the disagreement in Δ*c* between the two rotator positions is always smaller than the disagreement in Δ*c'*, as expected.

| Table 1 | | | |
|---|---|---|---|
| **crystals compared** | **spicule orientation** | **Δ*c'* (°)** | **Δ*c* (°)** |
| 1, 2 | Vertical | 3.9 | 4.1 |
| 1, 2 | Horizontal | 5.8 | 5.6 |
| 3, 4 | Vertical | 3.1 | 3.1 |
| 3, 4 | Horizontal | 5.1 | 3.3 |
| 5, 6 | Vertical | 1.6 | 1.7 |
| 5, 6 | Horizontal | 4.7 | 2.9 |
| 7, 8 | Vertical | 14.8 | 13.6 |
| 7, 8 | Horizontal | 20.2 | 16.2 |

*Table 1: Angular distances for crystals in the spicule of Figure 2, measured in different positions: vertical, or horizontal with the tip on the right. Crystals 1-8 are shown in Figure S2. The angular distance of two adjacent crystals is measured in two different ways: Δc' is the angular distance of the c'-axes, thus in 2D; Δc is the angular distance of the c-axes in 3D. Rotating the spicule from vertical to horizontal yields identical measurements, within an error of 3°, for both Δc' and Δc. Furthermore, Δc' and Δc are within 4° of one another. See SI sections 4.4 and 4.6 for further details, and Figure S3 for a schematic showing Δc' and Δc. In both measurements the uncertainty is 2°* [37].

The biogenic vaterite crystals in Figures 1 and 2 have small angle spreads, between 0° and 30°, as shown in Figure 3.

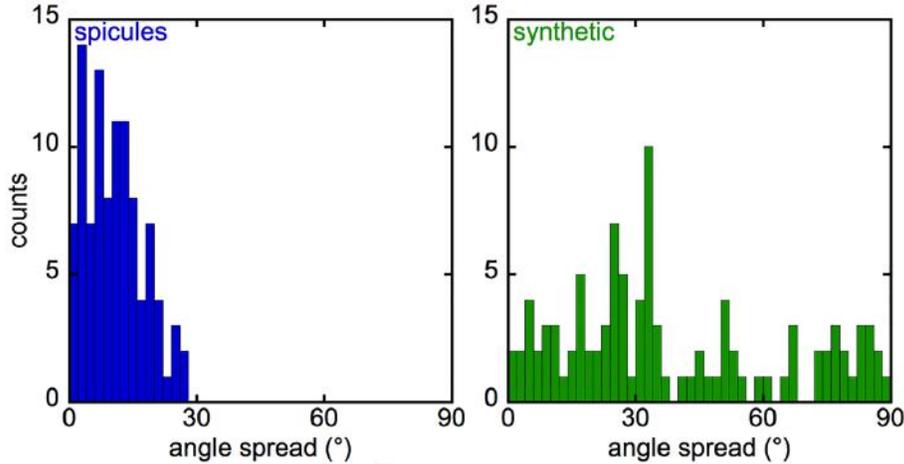

*Figure 3: Histogram of angular distances Δc' between two adjacent crystals in biogenic vaterite spicules and synthetic vaterite, measured from the PIC-maps in Figure 1. The biogenic vaterite crystals only show small angular distances (0°-30°), whereas synthetic vaterite domains of nanoparticles are randomly oriented (0°-90°).*

In Figure 3 we show the angular distance of *c'*-axes across the interface of vaterite orientation domains of nanoparticles in synthetic vaterite or single crystals in spicules. Clearly the orientation within a nanoparticle domain in synthetic vaterite is most often homogeneous but adjacent domains are randomly oriented ("S" in Figure 1, and Figure 3). On the other hand vaterite spicules consistently show much smaller angular distances, less than 30°.

In Figures 4, S4-S6 we present SEM micrographs of the core and thorns of body spicules cryo-fractured in liquid $N_2$ to expose their fracture figure, which is smooth and does not exhibit nanoparticles.

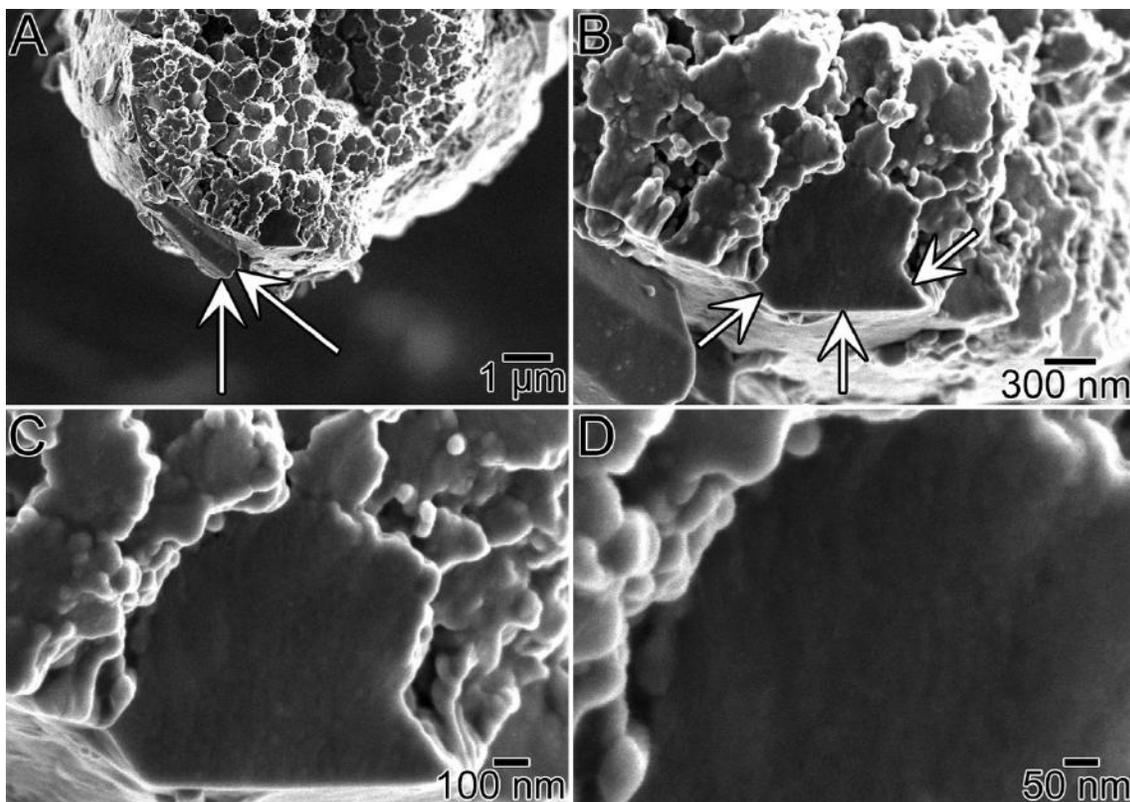

*Figure 4: SEM micrographs of a cryo-fractured body spicule. No nanoparticulate texture appears in a broken thorn. Two thorns show crystalline faces (arrows in A and B) of two euhedral (hexagonal pyramid) crystals, which are more clearly visible in B and C. Notice the smooth thorn fracture figure in D. Additional thorns and core crystals are shown in Figures S4-S6.*

Figure 5 shows that vaterite crystals with different orientations directly abut one another, with no organics at the interface.

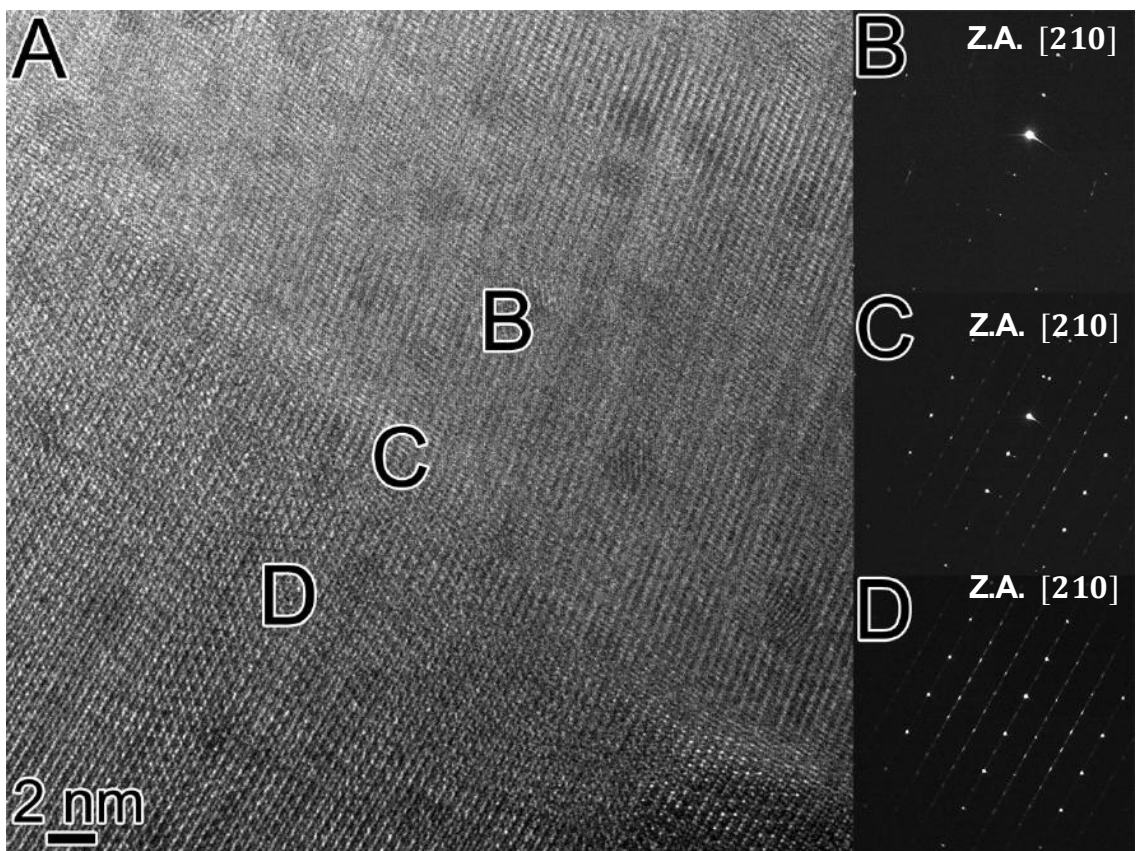

*Figure 5: A. HRTEM micrograph of a portion of body spicule, including a core and a thorn crystal. B, C, D. Nanobeam linescan electron diffraction obtained with 3-nm beam from the core (B), the interface (C), and the thorn crystal (D). The two crystal lattices differ in orientation, but there are no organics or other discontinuities at their interface. The interface clearly shows reflections from both crystal lattices. Additional abutting crystals are shown in Figures S7.*

| Technique | Parameter measured | Observation in spicules |
|---|---|---|
| PIC-mapping | Orientation of crystals, 20 nm resolution | Angle spreads <30° |
| FIB + TEM | Orientation of crystals, 2 nm resolution | Angle spreads <30°; No organics between adjacent crystals |
| SEM | Surface morphology | Faceted thorn crystals, smooth non-nanoparticulate cryo-fractured crystals |

*Table 2: Summary of the techniques used in this work, the parameters each measured, and the main observations they provided in vaterite spicules.*

**Discussion**
The data in Figures 1-3 show that individual crystallites in vaterite spicules form an elongated, structure with core crystals and euhedral thorns (Figure 4), all of which are space filling, with neither voids or organics at the interface of differently oriented crystals (Figure 5). These crystallites have narrowly distributed angle spreads, within 0°-30° around the spicule axis, thus all crystals in a spicule have their *c*-axes aligned with the spicule axis, not perfectly, but within ±15°.

Similar 0°-30° angle spreads were previously observed in nacre [20], where aragonite ($CaCO_3$) crystalline tablets are separated by organic sheets everywhere, except in Checa bridges [38, 39]. In materials science, interfaces between crystals with similar, but usually smaller mis-orientation angles, on the order of 15° and below, are termed "low angle grain boundaries" or "mosaicity" [40-42].

The observed angle spreads of 0°-30° make vaterite spicules different from single crystals, either bulk or branched as snowflakes, and from fractal crystal growth, where adjacent crystals are randomly oriented.
They likely correspond to low angle boundaries, which have been observed in many materials with high degree of crystallinity, including metals, ceramics, minerals, and molecular crystals [42]. The lack of organics at crystal interfaces, shown in Figure 5 and Figures S7, corroborate this interpretation.

Crystals growing with low-angle branching (e.g. 30°) were previously observed, and termed "non-crystallographic branching", in spherulitic crystals [43-45]. "Spherulitic crystal growth" is defined as radial polycrystalline growth resulting from successive non-crystallographic branching (NCB) from a central nucleus [43, 46]. The vaterite spicules presented here clearly do not resemble spheres, and do not exhibit the radially distributed acicular crystals that in spherulites appear at all angles, starting from a single center. However, locally the vaterite crystals in each spicule may be interpreted as a small-angle sector of a spherulite, in which the centers of radially distributed crystals move along the spicule axis as the spicule grows, making this a feather-like or "plumose" spherulite [47]. The present data do not demonstrate spherulitic growth, because low-angle branching of crystals is necessary but not sufficient to identify spherulitic growth. Hence other mechanisms, distinct from spherulitic growth, may produce nucleation of crystals with lattices at 0°-30° angle spreads, on the surface of previous crystals.

Adjacent crystals are always space-filling [48] and directly abut one another inside the spicule, at either end of each spicule and along its length. They were therefore never observed to form with the morphology of branches or dendrites, hence the term "branching" does not seem appropriate to describe crystal lattices tilted by a 0°-30° angle from one another.

Spherulitic growth is observed most frequently in crystals formed ion-by-ion from solution [43], in sulphates [49, 50] and oxalates [51], or via phase-transformation, for instance in the case of glass to crystalline phase, e.g. in small organic molecules [52, 53], or crystallization of polymers from the melt, e.g. in plastics [54] or metals [54, 55]. If vaterite spicules grow spherulitically, it is conceivable, therefore, that they either grow by attachment of amorphous precursor particles and solid-state transformation to crystalline vaterite, as indicated by the Wolf group in *in vitro* studies [56], or grow ion-by-ion from solution. We attempted to distinguish between these two possibilities with SEM experiments. In Figures 4, S4, S5 we noticed that the fracture figure of vaterite crystals is smooth and does not exhibit nanoparticles. The presence of nanoparticulate fracture demonstrates growth by particle attachment; its absence, however, does not rule it out. Crystal growth from solution is a possible interpretation of smooth crystal fracture.

Figure S8 shows a typical cryo-fracture figure of a sea urchin spicule, which does form by aggregation of amorphous precursor nanoparticles [57], and fractures accordingly [57, 58]. We only know the formation mechanisms of a few biominerals, but for all those cases, nanoparticulate fracture is a shared character [59, 60].

The smooth fracture figures of core crystallites and thorns in spicules suggest that these vaterite crystals may not have grown nanoparticle-by-nanoparticle [60] but ion-by-ion from solution, or by aggregation of particles followed by dissolution and re-precipitation.
Additional evidence consistent with the possibility of ion-by-ion crystal growth is provided by the crystal faces observed in vaterite spicule thorns (Figures 4 and S6). Such euhedral crystals, with flat faces and sharp corners, are rarely observed in mature, eukaryotic biominerals forming via amorphous precursors. They have been observed in biominerals formed by unknown mechanisms in unicellular organisms [61, 62], and in only two other animal biominerals: (i) the limpet radula teeth, which are made of goethite, show crystal faces common in synthetic and geologic goethite, and appear to form from solution [63]; and (ii) enamel hexagonal nano-rods, which form via an amorphous precursor at first [64], and then overgrowth is from solution [65]. Observation of euhedral crystals in vaterite spicules may therefore indicate that these crystals grew from solution. Again this is not conclusive proof, simply an indication, which in our opinion is stronger than the smooth fracture. Furthermore, the two separate and independent indications, smooth fracture and faceted crystal morphology, strengthen one another.

The ion-by-ion crystal growth suggested by Figures 4 and S4-S6 does not exclude that amorphous nanoparticles were *initially* delivered to the mineralization site. If such particles were present, they may have undergone dissolution and re-precipitation, thus only the *final stage* of crystal formation was via ion-by-ion growth from solution. This is the case in synthetic vaterite growth, which starts from amorphous calcium carbonate that rapidly dissolves and re-precipitates as vaterite [66].

Finally, Figure 5 shows that vaterite crystals with different orientations directly abut one another, with no organics at the interface. This is also unusual for biominerals, and typical of polycrystalline materials growing abiotically [67].

All these independent lines of evidence concur to demonstrate that vaterite spicules are most unusual among eukaryotic biominerals, and may point in the direction of possible growth mechanisms, which remain to be demonstrated. Such mechanisms are strongly constrained by the present observation of crystals nucleating within 0°-30° angle spreads and immediately abutting other crystals.

If the spicules grow from solution, we see, among others, three possible scenarios: (1) The whole spicule is one single crystal, resulting from one nucleation event, and the differently-oriented crystallites result from internal or external stress during the crystal growth process. (2) There is a new nucleation event for each differently-oriented crystallite, with similar but not identical orientation. (3) The similarity of orientation is a result of faster growth rate along the crystallographic *c*-axis and confinement in an organic compartment that does not allow the spicule to expand radially but only to grow longitudinally. The first scenario is consistent with the observation that spicules are extremely flexible when seen at the optical microscope in a droplet of ethanol as it evaporates and convects vigorously. If the spicule is bent during crystal growth, the stress could be significant, and the resulting growth strained and mis-oriented. The second and third are plausible, as in other biomineral similar mechanisms have been observed [19, 68, 69].

Perhaps the most promising avenue to pursue in elucidating such formation mechanisms is a broader study of spherulitic biominerals.

Many biominerals have been assumed from their morphology to be spherulitic, including corals [70], vertebrate otoconia [71], crustacean statoliths [72], fish otoliths [73, 74], and avian eggshells [75, 76]. Some corals show a radial distribution of crystal orientations in transmission PIC-mapping [77], others show random orientations in polarized light microscopy [78]. Some eggshells have randomly oriented calcite crystals, others a preferred radial orientation [79]. Only bio-induced, not bio-controlled [7], kidney stones have been demonstrated to form spherulitically, with crystal orientation analysis [80-83]. For all other biominerals, however, there is no high-resolution quantitative analysis showing the orientations of crystals and their cryo-fracture figures. In the absence of such data it is hard to assess whether the present results are widespread or rare.

Orientation analysis will demonstrate whether or not other biominerals grow spherulitically, and show differences or similarities with vaterite spicules. We stress that high-resolution PIC-mapping was necessary in order to measure the nano- and micro-crystal orientations described here in vaterite spicules. Coarser resolution, however, is sufficient for quantitative crystal orientation analysis of larger biominerals, including corals and eggshells, thus narrowly distributed angle spreads can be demonstrated using x-ray diffraction, or even simple visible light microscopy with crossed polarizers. Once sufficient studies of other biominerals are completed, the significance of spherulitic biomineral growth will be clearer.

In synthetic vaterite, the morphology of crystals depends on the growth conditions evolving from hexagonal monocrystalline plates, to florets and finally to spherulites as the super-saturation increases [84]. Adding alcohol also changes the morphology of synthetic vaterite crystals [44]. Unknown, biologically controlled conditions for biomineral formation, therefore, may determine the morphology and crystal orientation patterns in vaterite biominerals, such as the spicules described here or defective vaterite mineralization in mollusk shells [85-87], freshwater lackluster pearls [88], green turtle eggshells [89], and coho salmon otoliths [90].

Few biominerals have been studied with this question in mind, but it is possible that ion-by-ion growth from solution is a widespread growth mechanism, e.g. in poorly controlled biological mineralization processes such as in calcareous algae.

The evolutionary advantage, if there is one, of making vaterite spicules, instead of calcite or aragonite, remains obscure. Among the three anhydrous polymorphs of calcium carbonate, vaterite is the most soluble, has the lowest density, has no hydrated polymorphs, and is the least thermodynamically stable, but *H. momus* masters vaterite stabilization;

in fact, its spicules remain vateritic even years after extraction from the animal, or a year in seawater [23]. It is possible that there exist correlations between the vaterite polymorph selection, the large crystal sizes, the 0°-30° orientation angles described here, and the mechanical support function of the spicules. Future experiments will investigate possible correlations.

## Methods
Detailed methods are described in SI Detailed Methods. Briefly, vaterite spicules were extracted from *H. momus*, embedded in epoxy, polished, coated with 1 nm Pt in the area to be analyzed by PEEM and 40 nm around it, as described in refs. [31-33]. PEEM experiments were done on PEEM-3 [35] at the Advanced Light Source in Berkeley, CA, USA. SEM and TEM experiments were done at Technion, Haifa, Israel, using a Zeiss Ultra-Plus Field Emission Gun (FEG)-SEM, and an aberration-corrected Titan FEI (S)TEM.

Synthetic vaterite synthesis: 100 ml of 50 mM $CaCl_2 \cdot 2H_2O$ was equilibrated by KOH to have a pH of 13. In parallel, 100 ml of 50 mM $NaHCO_3$ solution was prepared. Both solutions were cooled to 5°C, after which the $CaCl_2$ solution was added to the second solution via a syringe pump at a rate of 1.5 ml/min, over 1 hour with gentle stirring. The formed powder was filtered and air-dried at room temperature followed by drying in a vacuum oven at 60°C for 2 hours.

## Supporting Information Available
Figures S1-S9, Detailed Methods. This information is available free of charge via the Internet at http://pubs.acs.org/.

## Acknowledgements
We thank Steve Weiner for his review of the manuscript and suggestions for improvements before submission, and Lara Estroff for discussions. BP and PG acknowledge joint support from US-Israel Binational Science Foundation (BSF-2010065). PG acknowledges support from NSF (DMR-1105167), DOE (DE-FG02-07ER15899), and the Radcliffe Institute for Advanced Study at Harvard University. BP acknowledges support from the European Research Council under the European Union's Seventh Framework Program (FP/2007–2013)/ERC Grant Agreement n° [336077]. PEEM experiments were done at the ALS, supported by DOE grant DE-AC02-05CH11231.